\documentclass[preprint,pre,showpacs,showkeys]{revtex4}
\usepackage{amssymb}
\usepackage{amsmath}
\usepackage{amsfonts}

\newtheorem{theorem}{Theorem}

\newenvironment{proof}[1][Proof]{\noindent\textbf{#1.} }{\ \rule{0.5em}{0.5em}}

\begin{document}

\title[]{Law of Error in Tsallis Statistics}
\author{Hiroki Suyari}
\affiliation{Department of Information and Image Sciences, Faculty of Engineering, Chiba
University, 263-8522, Japan}
\email{suyari@ieee.org, suyari@faculty.chiba-u.jp}
\author{Makoto Tsukada}
\affiliation{Department of Information Sciences, Faculty of Science, Toho University,
274-8510, Japan}
\email{tsukada@is.sci.toho-u.ac.jp}

\begin{abstract}
Gauss' law of error is generalized in Tsallis statistics such as
multifractal systems, in which Tsallis entropy plays an essential role
instead of Shannon entropy. For the generalization, we apply the new
multiplication operation determined by the $q-$logarithm and the $q-$%
exponential functions to the definition of the likelihood function in Gauss'
law of error. The maximum likelihood principle leads us to finding Tsallis
distribution as nonextensively generalization of Gaussian distribution.
\end{abstract}

\volumeyear{}
\volumenumber{}
\issuenumber{}
\eid{}
\date{\today }
\startpage{1}
\endpage{1}
\maketitle

\section{Introduction}

The maximum entropy principle for \textit{Tsallis entropy }\cite{Ts88}\cite%
{CT91} :%
\begin{equation}
S_{q}:=\frac{1-\int f\left( x\right) ^{q}dx}{q-1}\quad \left( q\in \mathbb{R}%
^{+}\right)  \label{Tsallis entropy}
\end{equation}%
under the constraints:%
\begin{equation}
\int f\left( x\right) dx=1,\quad \frac{\int x^{2}f\left( x\right) ^{q}dx}{%
\int f\left( y\right) ^{q}dy}=\sigma ^{2}
\end{equation}%
yields the so-called \textit{Tsallis distribution}:%
\begin{equation}
f\left( x\right) =\frac{\exp _{q}\left( -\beta _{q}x^{2}\right) }{\int \exp
_{q}\left( -\beta _{q}y^{2}\right) dy}\varpropto \left[ 1+\left( 1-q\right)
\left( -\beta _{q}x^{2}\right) \right] ^{\frac{1}{1-q}},
\label{Tsallis distribution}
\end{equation}%
where $\exp _{q}\left( x\right) $ is the $q-$\textit{exponential function}
defined by 
\begin{equation}
\exp _{q}\left( x\right) :=\left\{ 
\begin{array}{ll}
\left[ 1+\left( 1-q\right) x\right] ^{\frac{1}{1-q}} & \text{if }1+\left(
1-q\right) x>0, \\ 
0 & \text{otherwise}%
\end{array}%
\right. \quad \left( x\in \mathbb{R}\right)  \label{q-exponential}
\end{equation}%
and $\beta _{q}$ is a positive constant related to $\sigma $ and $q$ \cite%
{TLSM95}\cite{PT00}. For $q\rightarrow 1$, Tsallis distribution (\ref%
{Tsallis distribution}) recovers a Gaussian distribution. The power form in
Tsallis distribution (\ref{Tsallis distribution}) has been found to be
fairly fitted to many physical systems which \textit{cannot} be
systematically studied in the usual Boltzmann-Gibbs statistical mechanics 
\cite{AO01}.

The mathematical basis for Tsallis statistics comes from the deformed
expressions for the logarithm and the exponential functions which are the $%
q- $\textit{logarithm function}:%
\begin{equation}
\ln _{q}x:=\frac{x^{1-q}-1}{1-q}\quad \left( x\geq 0,q\in \mathbb{R}%
^{+}\right)  \label{q-logarithm}
\end{equation}%
and its inverse function, the $q-$\textit{exponential function }(\ref%
{q-exponential})\textit{.} Using the $q-$logarithm function (\ref%
{q-logarithm}), Tsallis entropy (\ref{Tsallis entropy}) can be written as%
\begin{equation}
S_{q}=-\int f\left( x\right) ^{q}\ln _{q}f\left( x\right) dx,
\end{equation}%
which is easily found to recover Shannon entropy when $q\rightarrow 1$.

The successful many applications of Tsallis statistics stimulate us to try
to find the new mathematical structure behind Tsallis statistics \cite{Su02a}%
. Some physicists reported some relations to the $q-$analysis which has been
usually discussed in quantum group \cite{Ab97}\cite{Jo98}. Apart from the
relations to the $q-$analysis, Borges presents a deformed algebra related to
the $q-$logarithm and the $q-$exponential function which naturally emerge
from Tsallis entropy \cite{Bo03}.

Using the algebra introduced by Borges, we derive Tsallis distribution by
applying the multiplication operation determined by the $q-$logarithm and
the $q-$exponential function to the \textit{maximum likelihood principle}
(MLP for short). Concretely, the new multiplication operation $\otimes _{q}$%
, which is called the $q-$\textit{product} in \cite{Bo03}, is applied to the 
\textit{likelihood function }in\textit{\ Gauss' law of error} with the
result that we obtain Tsallis distribution as nonextensively generalization
of Gaussian distribution. The present derivation of Tsallis distribution is
another way than the above maximum entropy principle (MEP for short, in
contrast to MLP).

\section{Gauss' law of error}

In order to clarify our ideas in the present work in section IV, we briefly
review Gauss' law of error in this section.

Consider the following situation: we get $n$ observed values:%
\begin{equation}
x_{1},x_{2},\cdots ,x_{n}\in \mathbb{R}  \label{data}
\end{equation}%
as results of mutually independent $n$ measurements for some observation.
Each observed value $x_{i}\,\left( i=1,\cdots ,n\right) $ is each value of
independent, identically distributed (i.i.d. for short) random variable $%
X_{i}\,\left( i=1,\cdots ,n\right) $, respectively. There exist a true value 
$x$ satisfying the \textit{additive} relation:%
\begin{equation}
x_{i}=x+e_{i}\quad \left( i=1,\cdots ,n\right) ,  \label{relation}
\end{equation}%
where each of $e_{i}$ is an error in each observation of a true value $x$.
Thus, for each $X_{i}$, there exists a random variable $E_{i}$ such that $%
X_{i}=x+E_{i}\,\left( i=1,\cdots ,n\right) $. Every $E_{i}$ has the same
probability density function $f$ which is differentiable, because $%
X_{1},\cdots ,X_{n}$ are i.i.d. (i.e., $E_{1},\cdots ,E_{n}$ are i.i.d.).
Let $L\left( \theta \right) $ be a function of a variable $\theta ,$ defined
by%
\begin{equation}
L\left( \theta \right) =L\left( x_{1},x_{2},\cdots ,x_{n};\theta \right)
:=f\left( x_{1}-\theta \right) f\left( x_{2}-\theta \right) \cdots f\left(
x_{n}-\theta \right) .  \label{likelihood function}
\end{equation}

\begin{theorem}
If the function $L\left( x_{1},x_{2},\cdots ,x_{n};\theta \right) $ of $%
\theta $ for any fixed $x_{1},x_{2},\cdots ,x_{n}$ takes the maximum at%
\begin{equation}
\theta =\theta ^{\ast }:=\frac{x_{1}+x_{2}+\cdots +x_{n}}{n},
\label{maximum value}
\end{equation}%
then the probability density function $f$ must be a Gaussian probability
density function:%
\begin{equation}
f\left( e\right) =\frac{1}{\sqrt{2\pi }\sigma }\exp \left\{ -\frac{e^{2}}{%
2\sigma ^{2}}\right\} .  \label{Gaussian}
\end{equation}
\end{theorem}

The above result goes by the name of \textquotedblleft \textit{Gauss' law of
error}\textquotedblright\ \cite{Ha98} which is often used as an assumption
in measurements in many scientific fields. In fact, on the basis of Gauss'
law of error some functions such as error function are often used to
estimate error rate in measurements.

Note that the assumption (\ref{relation}) means each error $e_{i}$ is 
\textit{additive} to the true value $x$. This assumption essentially
contributes the determination of a Gaussian probability density function as
seen in the proof below. The last assumption is a representation of MLP. In
MLP, the parameter $\theta \ $and the function $L\left( \theta \right) $ are
called \textit{population parameter }and\textit{\ likelihood function,}
respectively. The statement of the last assumption is rewritten in MLP as
follows: for the likelihood function $L\left( \theta \right) $ given by (\ref%
{likelihood function}), the maximum likelihood estimator is%
\begin{equation}
\overset{\wedge }{\theta }:=\frac{X_{1}+X_{2}+\cdots +X_{n}}{n}.  \label{MLE}
\end{equation}

Here we present the rigorous proof of \textit{Gauss' law of error.}

\begin{proof}
Taking the logarithm of the both side of the likelihood function $L\left(
\theta \right) $ in (\ref{likelihood function}) leads to%
\begin{equation}
\ln L\left( \theta \right) =\ln f\left( x_{1}-\theta \right) +\ln f\left(
x_{2}-\theta \right) +\cdots +\ln f\left( x_{n}-\theta \right) .
\label{logarithmLF}
\end{equation}%
Differentiating the above formula (\ref{logarithmLF}) with respect to $%
\theta $, we have%
\begin{equation}
\frac{L\prime \left( \theta \right) }{L\left( \theta \right) }=-\frac{%
f\prime \left( x_{1}-\theta \right) }{f\left( x_{1}-\theta \right) }-\cdots -%
\frac{f\prime \left( x_{n}-\theta \right) }{f\left( x_{n}-\theta \right) }.
\label{bibun LLF}
\end{equation}%
The last assumption implies that when $\theta =\theta ^{\ast }$ the
likelihood function $L\left( \theta \right) $ takes the maximum value, so
that%
\begin{equation}
\frac{L\prime \left( \theta ^{\ast }\right) }{L\left( \theta ^{\ast }\right) 
}=0
\end{equation}%
i.e., 
\begin{equation}
\frac{f\prime \left( x_{1}-\theta ^{\ast }\right) }{f\left( x_{1}-\theta
^{\ast }\right) }+\cdots +\frac{f\prime \left( x_{n}-\theta ^{\ast }\right) 
}{f\left( x_{n}-\theta ^{\ast }\right) }=0.  \label{LLF=0}
\end{equation}%
Let $e_{i}^{\ast }$ be defined by%
\begin{equation}
e_{i}^{\ast }:=x_{i}-\theta ^{\ast }\quad \left( i=1,\cdots ,n\right) ,
\label{def e_i}
\end{equation}%
then (\ref{LLF=0}) can be rewritten to%
\begin{equation}
\sum_{i=1}^{n}\frac{f\prime \left( e_{i}^{\ast }\right) }{f\left(
e_{i}^{\ast }\right) }=0.  \label{LLF=02}
\end{equation}%
Using the new function $\phi \left( e\right) $ defined by%
\begin{equation}
\phi \left( e\right) :=\frac{f\prime \left( e\right) }{f\left( e\right) },
\end{equation}%
(\ref{LLF=02}) becomes%
\begin{equation}
\sum_{i=1}^{n}\phi \left( e_{i}^{\ast }\right) =0.  \label{condition1}
\end{equation}%
On the other hand, summing up the both sides of (\ref{def e_i}) for all $%
i=1,\cdots ,n$, we obtain%
\begin{equation}
\sum\limits_{i=1}^{n}e_{i}^{\ast }=\sum\limits_{i=1}^{n}x_{i}-n\theta ^{\ast
}=0,  \label{condition2}
\end{equation}%
where we used (\ref{maximum value}).

Then our problem can be reduced to finding the function $\phi$ satisfying (%
\ref{condition1}) under the constraint (\ref{condition2}). For simplicity,
\textquotedblleft*\textquotedblright\ in $e_{i}^{\ast}$\ is abbreviated in
the rest of this proof.

From (\ref{condition2}) we obtain 
\begin{equation}
\frac{de_{i}}{de_{1}}=-1\quad\left( i\neq1\right) ,  \label{e_i no bibun}
\end{equation}
so that differentiating (\ref{condition1}) with respect to $e_{1}$ goes to%
\begin{equation}
\frac{d\phi\left( e_{1}\right) }{de_{1}}-\frac{d\phi\left( e_{i}\right) }{%
de_{i}}=0\quad\left( i\neq1\right) .  \label{phi no bibun}
\end{equation}
The choice of $e_{1}$ does not lose the lack of generality in the above two
formulas (\ref{e_i no bibun}) and (\ref{phi no bibun}), so in general there
exists a constant $a\in\mathbb{R}$ such that%
\begin{equation}
\frac{d\phi\left( e\right) }{de}=a.
\end{equation}
Thus, it implies%
\begin{equation}
\phi\left( e\right) =ae+b  \label{phi}
\end{equation}
for some $b\in\mathbb{R}.$ Applying this formula (\ref{phi}) and (\ref%
{condition2}) to (\ref{condition1}), we get 
\begin{equation}
0=\sum_{i=1}^{n}\phi\left( e_{i}\right) =a\sum_{i=1}^{n}e_{i}+nb=nb,
\end{equation}
so it implies 
\begin{equation}
b=0.
\end{equation}
Therefore, 
\begin{equation}
\phi\left( e\right) =ae\quad\Leftrightarrow\quad\frac{f\prime\left( e\right) 
}{f\left( e\right) }=ae.  \label{bibun_houteishiki}
\end{equation}
The solution $f$ satisfying (\ref{bibun_houteishiki}) can be obtained to%
\begin{equation}
f\left( e\right) =C\exp\left( \frac{ae^{2}}{2}\right)
\end{equation}
where $C$ is a constant. $f$ is an even function and a probability density
function, so that $a<0$. Thus, there exists a constant $h>0$ such that $a$
can be set to%
\begin{equation}
\frac{a}{2}=-h^{2}.
\end{equation}
Then, $f$ becomes%
\begin{equation}
f\left( e\right) =\frac{h}{\sqrt{\pi}}\exp\left( -h^{2}e^{2}\right)
\end{equation}
where $\int f\left( e\right) de=1$ is used. $h$ can defined by 
\begin{equation}
h:=\frac{1}{\sqrt{2}\sigma}>0,
\end{equation}
therefore we can determine $f$ as a Gaussian probability density function
with mean zero and variance $\sigma^{2}$:%
\begin{equation}
f\left( e\right) =\frac{1}{\sqrt{2\pi}\sigma}\exp\left\{ -\frac{e^{2}}{%
2\sigma^{2}}\right\} .
\end{equation}
\end{proof}

Note that Gauss' law of error tells us that in measurements it is the most
probable to assume a Gaussian probability distribution for \textit{additive}
noise, which is often applied to every scientific fields.

In section IV, Gauss' law of error is generalized to the nonextensive
systems, which results in Tsallis distribution as a generalization of
Gaussian distribution.

\section{The new multiplication operation determined by\newline
\noindent$q-$logarithm function and $q-$exponential function}

The new multiplication operation $\otimes _{q}$ is first introduced by
Borges in \cite{Bo03} for satisfying the following equations:%
\begin{align}
\ln _{q}\left( x\otimes _{q}y\right) & =\ln _{q}x+\ln _{q}y,
\label{property of ln_q} \\
\exp _{q}\left( x\right) \otimes _{q}\exp _{q}\left( y\right) & =\exp
_{q}\left( x+y\right) .
\end{align}%
These lead us to the definition of $\otimes _{q}$ between two positive
numbers%
\begin{equation}
x\otimes _{q}y:=\left\{ 
\begin{array}{ll}
\left[ x^{1-q}+y^{1-q}-1\right] ^{\frac{1}{1-q}}, & \text{if }%
x>0,\,y>0,\,x^{1-q}+y^{1-q}-1>0 \\ 
0, & \text{otherwise}%
\end{array}%
\right. ,  \label{def of q-product}
\end{equation}%
which is called $q-$\textit{product} in \cite{Bo03}. The $q-$\textit{product}
recovers the usual product such that $\underset{q\rightarrow 1}{\lim }\left(
x\otimes _{q}y\right) =xy$. The fundamental properties of the $q-$product $%
\otimes _{q}$ are almost the same as the usual product, but the distributive
law does not hold in general.%
\begin{equation}
a\left( x\otimes _{q}y\right) \neq ax\otimes _{q}y\quad \left( a,x,y\in 
\mathbb{R}\right)
\end{equation}%
The properties of the $q-$\textit{product} can be found in \cite{Bo03}. For
the purpose of this paper, the differentiation of the $q-$logarithm function
is also needed in the next section.%
\begin{equation}
\frac{d}{dx}\ln _{q}x=\frac{1}{x^{q}}  \label{bibun of ln_q}
\end{equation}

\section{Nonextensively generalization of Gauss' law of error}

This section presents the law of error in Tsallis statistics with the
rigorous proof along the same line as that of Gauss' law of error.

Consider the same setting as $n$ observed values, as Gauss' law of error: we
get $n$ observed values:%
\begin{equation}
x_{1},x_{2},\cdots ,x_{n}\in \mathbb{R}
\end{equation}%
as results of $n$ measurements for some observation. Each observed value $%
x_{i}\,\left( i=1,\cdots ,n\right) $ is each value of identically
distributed random variable $X_{i}\,\left( i=1,\cdots ,n\right) $,
respectively. There exist a true value $x$ satisfying the \textit{additive}
relation:%
\begin{equation}
x_{i}=x+e_{i}\quad \left( i=1,\cdots ,n\right) ,  \label{q-relation}
\end{equation}%
where each of $e_{i}$ is an error in each observation of a true value $x$.
Thus, for each $X_{i}$, there exists a random variable $E_{i}$ such that $%
X_{i}=x+E_{i}\,\left( i=1,\cdots ,n\right) $. Every $E_{i}$ has the same
probability density function $f$ which is differentiable, because $%
X_{1},\cdots ,X_{n}$ are i.i.d. (i.e., $E_{1},\cdots ,E_{n}$ are i.i.d.).
Let $L_{q}\left( \theta \right) $ be a function of a variable $\theta ,$
defined by%
\begin{equation}
L_{q}\left( \theta \right) =L_{q}\left( x_{1},x_{2},\cdots ,x_{n};\theta
\right) :=f\left( x_{1}-\theta \right) \otimes _{q}f\left( x_{2}-\theta
\right) \otimes _{q}\cdots \otimes _{q}f\left( x_{n}-\theta \right) .
\label{q-likelihood function}
\end{equation}

\begin{theorem}
If the function $L_{q}\left( x_{1},x_{2},\cdots ,x_{n};\theta \right) $ of $%
\theta $ for any fixed $x_{1},x_{2},\cdots ,x_{n}$ takes the maximum at%
\begin{equation}
\theta =\theta ^{\ast }:=\frac{x_{1}+x_{2}+\cdots +x_{n}}{n},
\label{q-maximum value}
\end{equation}%
then the probability density function $f$ must be a Tsallis distribution:%
\begin{equation}
f\left( e\right) =\frac{\exp _{q}\left( -\beta _{q}e^{2}\right) }{\int \exp
_{q}\left( -\beta _{q}e^{2}\right) de}  \label{Tsallis}
\end{equation}%
where $\beta _{q}$ is a $q-$dependent positive constant.
\end{theorem}

\begin{proof}
Taking the $q-$logarithm of the both side of the likelihood function $%
L_{q}\left( \theta \right) $ in (\ref{q-likelihood function}) leads to%
\begin{equation}
\ln _{q}L_{q}\left( \theta \right) =\ln _{q}f\left( x_{1}-\theta \right)
+\ln _{q}f\left( x_{2}-\theta \right) +\cdots +\ln _{q}f\left( x_{n}-\theta
\right)   \label{q-logarithmLF}
\end{equation}%
where the property (\ref{property of ln_q}) is used. Differentiating the
above formula (\ref{q-logarithmLF}) with respect to $\theta $ according to (%
\ref{bibun of ln_q}), we have%
\begin{equation}
\frac{L_{q}\prime \left( \theta \right) }{\left( L_{q}\left( \theta \right)
\right) ^{q}}=-\frac{f\prime \left( x_{1}-\theta \right) }{\left( f\left(
x_{1}-\theta \right) \right) ^{q}}-\cdots -\frac{f\prime \left( x_{n}-\theta
\right) }{\left( f\left( x_{n}-\theta \right) \right) ^{q}}.
\label{q-bibun LLF}
\end{equation}%
When $\theta =\theta ^{\ast }$ the likelihood function $L_{q}\left( \theta
\right) $ takes the maximum, so that%
\begin{equation}
\frac{L_{q}\prime \left( \theta ^{\ast }\right) }{\left( L_{q}\left( \theta
^{\ast }\right) \right) ^{q}}=0\quad \Leftrightarrow \quad \frac{f\prime
\left( x_{1}-\theta ^{\ast }\right) }{\left( f\left( x_{1}-\theta ^{\ast
}\right) \right) ^{q}}+\cdots +\frac{f\prime \left( x_{n}-\theta ^{\ast
}\right) }{\left( f\left( x_{n}-\theta ^{\ast }\right) \right) ^{q}}=0.
\label{q-LLF=0}
\end{equation}%
Let $e_{i}^{\ast }$ and $\phi _{q}\left( e\right) $ be defined by%
\begin{equation}
e_{i}^{\ast }:=x_{i}-\theta ^{\ast }\,\,\left( i=1,\cdots ,n\right) ,\quad
\phi _{q}\left( e\right) :=\frac{f\prime \left( e\right) }{\left( f\left(
e\right) \right) ^{q}},  \label{q-def e_i}
\end{equation}%
then (\ref{q-LLF=0}) can be rewritten to%
\begin{equation}
\sum_{i=1}^{n}\phi _{q}\left( e_{i}^{\ast }\right) =0.  \label{q-condition1}
\end{equation}%
On the other hand, summing up the both sides of the former formula in (\ref%
{q-def e_i}) for all $i=1,\cdots ,n$, we obtain%
\begin{equation}
\sum_{i=1}^{n}e_{i}^{\ast }=\sum_{i=1}^{n}x_{i}-n\theta ^{\ast }=0,
\label{q-condition2}
\end{equation}%
where we used (\ref{q-maximum value}).

Then our problem can be reduced to finding the function $\phi _{q}$
satisfying (\ref{q-condition1}) under the constraint (\ref{q-condition2}).
For simplicity, \textquotedblleft *\textquotedblright\ in $e_{i}^{\ast }$\
is abbreviated in the rest of this proof.

From (\ref{q-condition2}) we obtain 
\begin{equation}
\frac{de_{i}}{de_{1}}=-1\quad \left( i\neq 1\right) ,  \label{q-e_i no bibun}
\end{equation}%
so that differentiating (\ref{q-condition1}) with respect to $e_{1}$ goes to%
\begin{equation}
\frac{d\phi _{q}\left( e_{1}\right) }{de_{1}}-\frac{d\phi _{q}\left(
e_{i}\right) }{de_{i}}=0\quad \left( i\neq 1\right) .  \label{q-phi no bibun}
\end{equation}%
The choice of $e_{1}$ does not lose the lack of generality in the above two
formulas (\ref{q-e_i no bibun}) and (\ref{q-phi no bibun}), so in general
there exists a $q-$dependent constant $a_{q}\in \mathbb{R}$ such that%
\begin{equation}
\frac{d\phi _{q}\left( e\right) }{de}=a_{q}\quad \Leftrightarrow \quad \phi
_{q}\left( e\right) =a_{q}e+b_{q}  \label{q-phi}
\end{equation}%
for some $b_{q}\in \mathbb{R}.$ Applying this formula (\ref{q-phi}) and (\ref%
{q-condition2}) to (\ref{q-condition1}), we get 
\begin{equation}
0=\sum_{i=1}^{n}\phi _{q}\left( e_{i}\right)
=a_{q}\sum_{i=1}^{n}e_{i}+nb_{q}=nb_{q}\quad \Rightarrow \quad b_{q}=0.
\end{equation}%
Therefore, 
\begin{equation}
\phi _{q}\left( e\right) =a_{q}e\quad \Leftrightarrow \quad \frac{f\prime
\left( e\right) }{\left( f\left( e\right) \right) ^{q}}=a_{q}e.
\label{q-bibun_houteishiki}
\end{equation}%
Using the property (\ref{bibun of ln_q}), (\ref{q-bibun_houteishiki}) can be
integrated with respect to $e$:%
\begin{equation}
\ln _{q}f\left( e\right) =\frac{a_{q}e^{2}}{2}+C_{q}
\end{equation}%
where $C_{q}$ is a $q-$dependent constant. Thus we obtain%
\begin{equation}
f\left( e\right) ^{1-q}=1+\left( 1-q\right) \left( \frac{a_{q}e^{2}}{2}%
+C_{q}\right) .
\end{equation}%
When $1+\left( 1-q\right) \left( \frac{a_{q}e^{2}}{2}+C_{q}\right) >0$ and $%
1+\left( 1-q\right) C_{q}>0$, $f\left( e\right) $ can be rewritten as
follows:%
\begin{align}
f\left( e\right) & =\left[ 1+\left( 1-q\right) \left( \frac{a_{q}e^{2}}{2}%
+C_{q}\right) \right] ^{\frac{1}{1-q}} \\
& =\left[ 1+\left( 1-q\right) C_{q}\right] ^{\frac{1}{1-q}}\left[ 1+\left(
1-q\right) \frac{a_{q}}{2\left( 1+\left( 1-q\right) C_{q}\right) }e^{2}%
\right] ^{\frac{1}{1-q}} \\
& =\exp _{q}\left( C_{q}\right) \exp _{q}\left( -\beta _{q}e^{2}\right)
\varpropto \exp _{q}\left( -\beta _{q}e^{2}\right) =\left[ 1+\left(
1-q\right) \left( -\beta _{q}e^{2}\right) \right] ^{\frac{1}{1-q}}
\end{align}%
where%
\begin{equation}
\beta _{q}:=\frac{-a_{q}}{2\left( 1+\left( 1-q\right) C_{q}\right) }.
\end{equation}%
Thus, the normalization condition $\int f\left( e\right) de=1$ implies that 
\begin{equation}
\exp _{q}\left( C_{q}\right) =\frac{1}{\int \exp _{q}\left( -\beta
_{q}e^{2}\right) de}.  \label{nomalization}
\end{equation}

Moreover, since $L_{q}\left( \theta \right) $ takes the maximum at $\theta
=\theta ^{\ast }$, $\left. \frac{\partial ^{2}\ln _{q}L_{q}\left( \theta
\right) }{\partial \theta ^{2}}\right\vert _{\theta =\theta ^{\ast }}$ can
be computed as 
\begin{equation}
0>\left. \frac{\partial ^{2}\ln _{q}L_{q}\left( \theta \right) }{\partial
\theta ^{2}}\right\vert _{\theta =\theta ^{\ast }}=\frac{2n\left( -\beta
_{q}\right) }{\left( \int \exp _{q}\left( -\beta _{q}e^{2}\right) de\right)
^{1-q}},
\end{equation}%
so that $\beta _{q}>0$. Therefore, we can determine (\ref{Tsallis}) for the
probability density function $f$.
\end{proof}

Our result $f\left( e\right) $ in (\ref{Tsallis}) coincides with (\ref%
{Tsallis distribution}) by applying the $q-$product to MLP instead of MEP.

\section{Conclusion}

We obtain law of error in Tsallis statistics by applying the $q-$product to
MLP. The $q-$product is defined as the new multiplication operation derived
from the $q-$logarithm and $q-$exponential function. These functions, $q-$%
logarithm and $q-$exponential functions, naturally emerge from Tsallis
entropy. Our present result reveals that in Tsallis statistics the $q-$%
product can constitute some important probabilistic formulations. The model
for Tsallis distribution has been presented in \cite{CLDO01}. Moreover,
based on our result, the \textit{error functions} in Tsallis statistics can
be also formulated as similarly as those of Gauss' law of error.

\end{document}